\documentclass[aps,onecolumn,11pt,floatfix,altaffilletter,preprintnumbers,tightenlines,showpacs,showkeys,notitlepage,nofootinbib,superscriptaddress]{revtex4-1}
\pdfoutput=1
\usepackage[colorlinks=true,citecolor=blue,linkcolor=blue]{hyperref}
\usepackage[utf8]{inputenc}
\usepackage[T1]{fontenc}            
\usepackage[normalem]{ulem}
\usepackage{amsmath,amssymb}
\usepackage{mathtools}
\usepackage{epsfig}
\usepackage{graphicx}               
\usepackage{url}
\usepackage{color}
\usepackage{multirow}
\usepackage{placeins}
\usepackage[dvipsnames]{xcolor}
\usepackage{epstopdf}
\usepackage[shortlabels]{enumitem}
\usepackage[capitalize]{cleveref}
\usepackage{lipsum}
\usepackage{siunitx}
\usepackage{booktabs} 
\usepackage{titlesec} 
\usepackage{subfigure}
\usepackage[subfigure,titles]{tocloft} 
 
\allowdisplaybreaks

\makeatletter

\renewcommand{\p@subsection}{}
\makeatother

\titleformat*{\section}{\centering\bfseries\scshape}
\titlelabel{\thetitle\quad}
\titleformat*{\paragraph}{\bfseries}
\titlespacing*{\paragraph}{0pt}{3.25ex plus 1ex minus .2ex}{1em}

\usepackage{xspace}

\pacs{}
\keywords{}

\begin{document}

\title{Timing and Multi-Channel:\\ Novel Method for Determining the Neutrino Mass Ordering from Supernovae}

\author{Vedran Brdar}
\email{vbrdar@fnal.gov}
\affiliation{Fermi National Accelerator Laboratory,
             Batavia, IL 60510, USA}
\affiliation{Northwestern University, Dept.~of Physics \& Astronomy,
             Evanston, IL 60208, USA}

\author{Xun-Jie Xu}
\email{ xuxj@ihep.ac.cn}
\affiliation{Institute of High Energy Physics, Chinese Academy of Sciences, Beijing 100049, China}

\preprint{nuhep-th/22-03, \,\,FERMILAB-PUB-22-364-T}

\begin{abstract}
One of the few remaining unknowns in the standard three-flavor neutrino oscillation paradigm is the ordering of neutrino masses. In this work we propose a novel method for determining neutrino mass ordering using the time information on early supernova neutrino events. In a core-collapse supernova, neutrinos are produced earlier than antineutrinos and, depending on the mass ordering which affects the adiabatic flavor evolution, may cause earlier observable signals in $\nu_e$ detection channels than in others. Hence, the time differences are sensitive to the mass ordering. We find that using the time information on the detection of the first galactic supernova events at future detectors like DUNE, JUNO and Hyper-Kamiokande, the mass ordering can already be determined at $\sim 2 \sigma$ CL, while $\mathcal{O}(10)$ events suffice for the discovery. Our method does not require high statistics and could be used within the supernova early warning system (SNEWS) which will have access to the time information on early supernova neutrino events recorded in a number of detectors. The method proposed in this paper also implies a crucial interplay between the mass ordering and the triangulation method for locating supernovae. 
\end{abstract}

\maketitle

\section{Introduction}
\label{sec:intro}
\noindent
The discovery of neutrino oscillations \cite{Super-Kamiokande:1998kpq,SNO:2002tuh,KamLAND:2002uet}, as one of the most relevant scientific achievements in the last few decades, has undoubtedly established that neutrinos are massive particles. Neutrino oscillation probabilities are functions of neutrino mass squared differences
which are precisely known by now \cite{Esteban:2020cvm}. However, the ordering of these masses is still unknown; namely, it is not yet established whether the neutrino mass eigenstate which comprises the largest fraction of electron neutrino flavor is the lightest one. If so, 
the neutrino mass ordering is dubbed ``normal'' (NO), while the alternative option is the ``inverted'' ordering (IO). 

There are several multi-billion dollar experiments that will start operating by the end of the decade, namely DUNE \cite{DUNE:2016hlj}, Hyper-Kamionade \cite{Hyper-Kamiokande:2018ofw} and JUNO \cite{JUNO:2015zny} and they are all expected
to have the capability to probe neutrino mass ordering \cite{DUNE:2016hlj,Hyper-Kamiokande:2022smq,Forero:2021lax}. As an alternative to man-made neutrinos from reactor and acceleration facilities, neutrinos from core-collapse supernove (SNe) can also be employed for the determination of mass ordering. Unlike for the case of SN 1987A \cite{Panagia:2004ii} from which only a handful of neutrinos were detected \cite{Kamiokande-II:1987idp}, the above mentioned experiments will be able to collect thousands of events from the next galactic supernova via several different interaction channels. 

Neutrinos from supernovae were already suggested in the context of neutrino mass ordering determination \cite{Dighe:1999bi,Kachelriess:2004ds,Scholberg:2017czd}.
One of the methods previously discussed was to focus on  neutrinos produced in the neutronization burst and to use event counts from the interaction of electron neutrinos ($\nu_e$) 
in detectors such as DUNE that excel at $\nu_e$ detection.
Due to matter effects in SNe \cite{Wolfenstein,Mikheev:1986gs,Mikheev:1986wj}, 
the survival probability of $\nu_e$ is higher and leads to more $\nu_e$ events in IO when compared to the NO scenario. 
Along similar lines, following the neutronization burst, electron antineutrinos ($\overline{\nu}_e$) will start to be produced in significant amounts 
and the matter effects in this case also affect the flux of $\overline{\nu}_e$ \cite{Lunardini:2003eh,Serpico:2011ir}, making NO and IO scenarios distinguishable in experiments like Super/Hyper-Kamiokande and JUNO that utilize inverse beta decay (IBD). Other interesting approaches include consideration of matter effects in Earth which lead to potentially detectable spectral distortions \cite{Lunardini:2001pb,Borriello:2012zc}, or using differences in arrival times of neutrino mass eigenstates~\cite{Jia:2017oar}. These approaches require either sufficiently high energy resolution or tremendously large-scale detectors. For a more detailed discussion on previously proposed techniques we refer the interested reader to the review \cite{Scholberg:2017czd}.

In this paper we propose a novel method for determining the mass ordering based on 
the time differences between neutrino and antineutrino events from a galactic SN.
Since early SN neutrinos are produced via neutronization, neutrinos are emitted from SN earlier than antineutrinos, leading to an observable time difference between  neutrino and antineutrino detection channels. Initially being produced from neutronization as $\nu_e$, the early neutrinos can be partially (IO) or almost entirely (NO) converted to $\nu_\mu$ and $\nu_\tau$ due to the adiabatic flavor evolution in the presence of matter effects~\cite{Dighe:1999bi}. 
Therefore, for $\nu_e$ events at e.g.~DUNE, the first few events in the NO are delayed when comparing to the IO scenario. For $\overline{\nu}_e$, which can be well measured via the IBD process at JUNO and Super/Hyper-Kamiokande, the timing of the first few events is less affected by the mass ordering. Hence, the time difference between the onset of events at DUNE and IBD detectors, after subtracting corrections due to different geographical locations of experiments,  provides a handle for the determination of the mass ordering\,\textemdash\,the larger it is, the likelihood for IO increases.  Our method utilizes electron neutrinos produced already during the infall phase of SN. Although the statistics of events induced by such neutrinos is low, it is well known that  neutrino fluxes in this period are practically model independent, making any mass ordering statement quite robust. As we will show, even by comparing the time difference between the first $\nu_e$ and $\overline{\nu}_e$ events at DUNE and JUNO respectively, one can already determine the mass ordering at $\sim 2 \sigma$ CL while only a handful of events are required to guarantee a discovery.

The method proposed in this paper can be particularly useful when incorporated in the supernova early warning system (SNEWS) \cite{SNEWS:2020tbu} which should have access to the first few neutrino events from all involved detectors well before the full data set of each experiment becomes available. This means that besides not requiring high statistics and being practically independent on the SN properties such as the progenitor mass, this method should be extremely time-efficient. In addition, we will show that this study is also helpful in the context of the  triangulation method~\cite{Beacom:1998fj,Brdar:2018zds,Linzer:2019swe} that can be utilized for determining a SN location via inter-detector time differences.

The paper is organized as follows. In \cref{sec:methods} we discuss SN fluxes for relevant (anti)neutrino flavors and compute respective event rates at various detectors.  In \cref{sec:results} we first discuss statistical methods employed for assessing the timing of neutrino interactions and calculate expected time window for the occurrence of particular events. This allows us to compute the statistical significance for the 
discrimination between NO and IO for the method proposed in this paper. In \cref{sec:triangulation} we discuss the non-trivial impact of the mass ordering on techniques proposed for determining SN location via triangulation. Finally,  we conclude in \cref{sec:conclusion}.

\section{Neutrino Event Rates and their dependence on the mass ordering }
\label{sec:methods}
\noindent

Neutrino fluxes from SNe have been calculated from extensive simulations, see e.g. Refs.~\cite{Hudepohl2013,Mirizzi:2015eza,Adam1,Adam2,Adam3,Burrows:2020qrp, Nakazato_2013,SNEWS:2021ezc}. 
In the main text, we will adopt the neutrino fluxes calculated by the Garching group~\cite{Hudepohl2013,Mirizzi:2015eza}.  We note, however, that neutrino fluxes produced by different groups are not fully consistent with each other. In order to investigate how this might affect our results, in \cref{sec:pr} we present results using fluxes from the Princeton group~\cite{Adam1,Adam2,Adam3, Burrows:2020qrp}. Qualitatively, our main conclusions are not changed by this difference.

\begin{figure}
  \centering
 \begin{tabular}{cc}
  \includegraphics[width=0.49\textwidth]{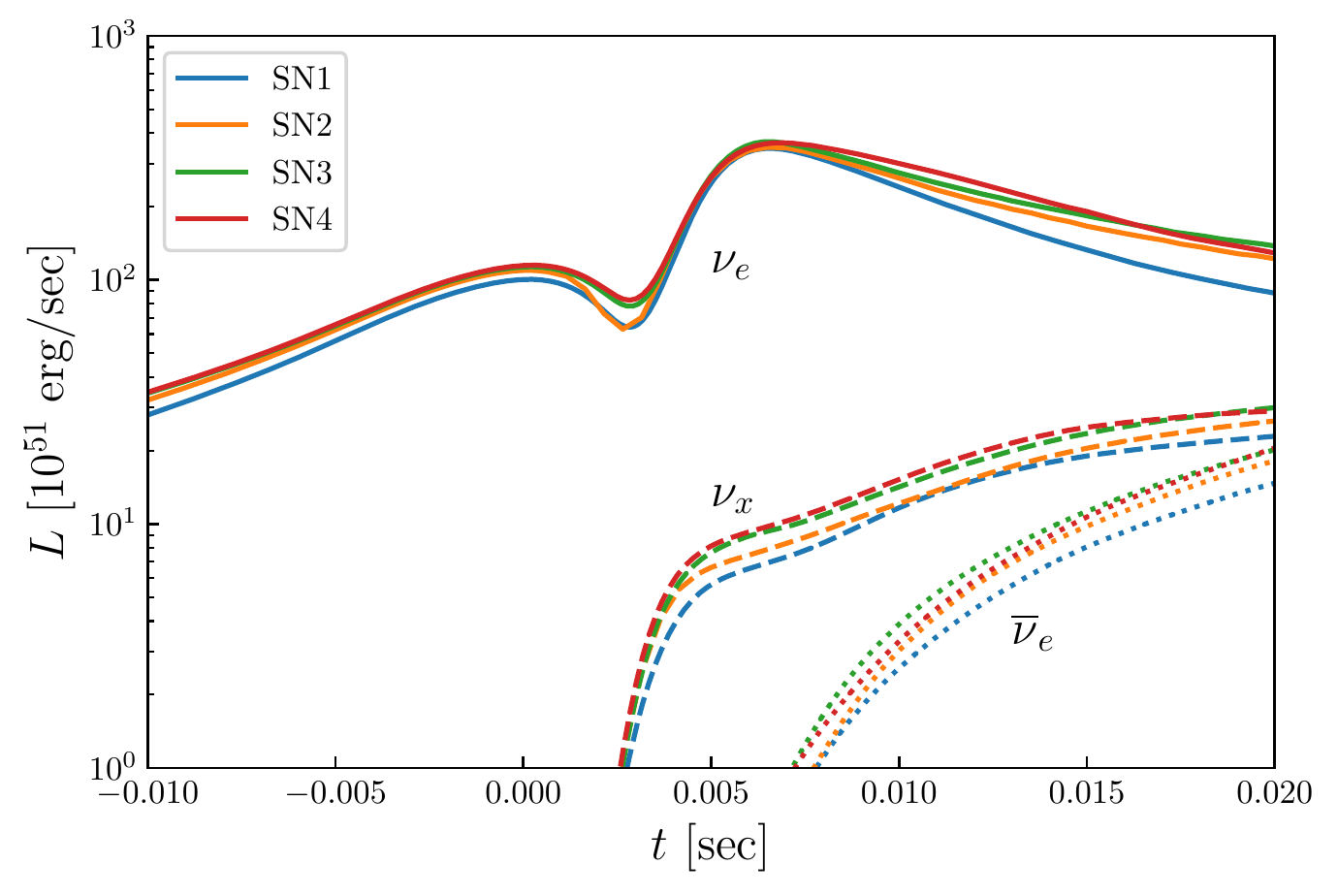} \includegraphics[width=0.48\textwidth]{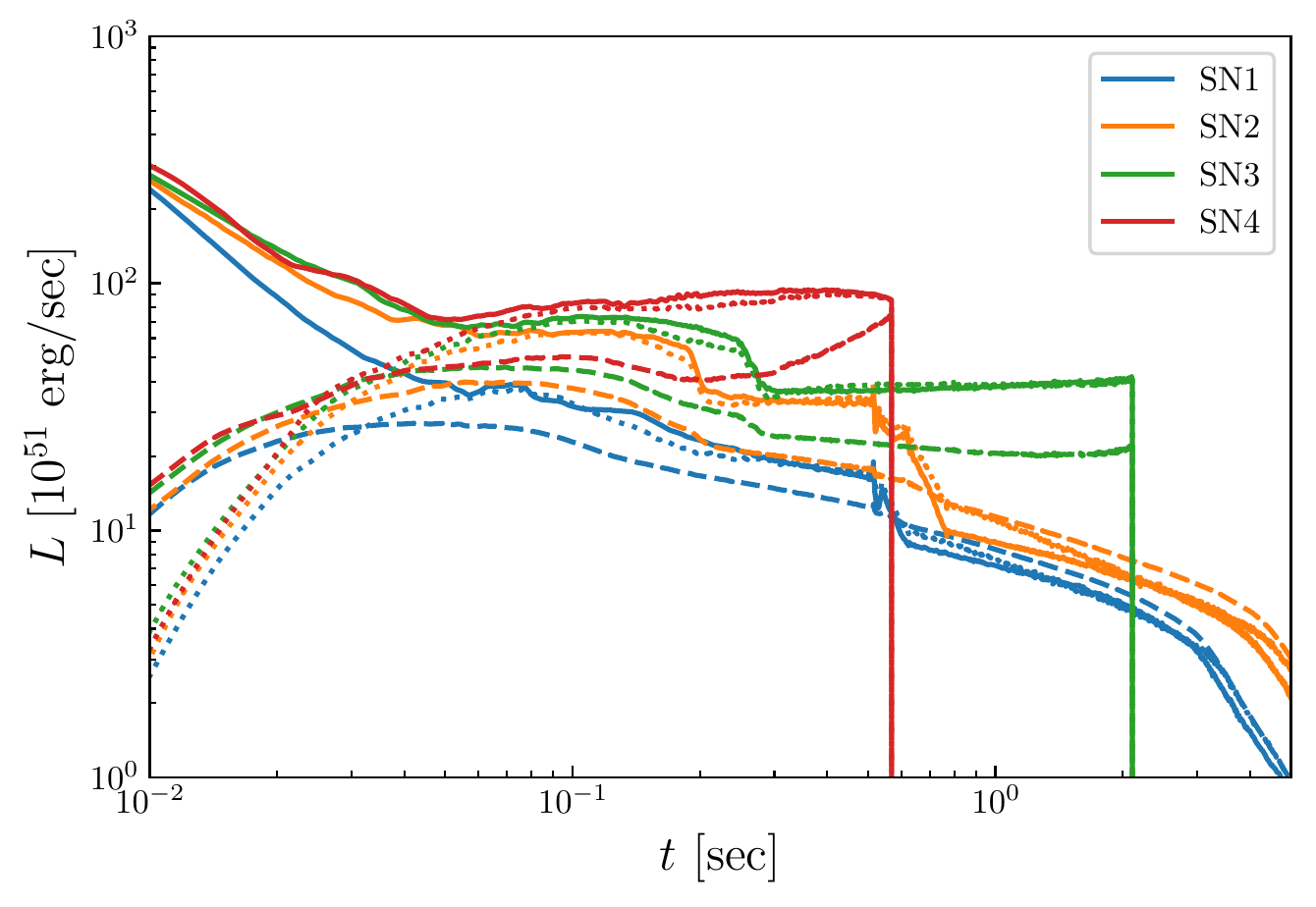} 
  \end{tabular}
  \caption{
    Neutrino luminosity curves for four considered SN models \cite{Hudepohl2013,Mirizzi:2015eza}. The left panel shows the period of infall and neutronization burst ($t<20$ ms after the core bounce) and in this time window neutrino luminosities are rather model-independent. In contrast, the luminosities for particular flavor associated to the accretion and cooling stage of SN (right panel) feature larger discrepancies across the considered models.   
\label{fig:L-t}
  }
  \label{fig:fluxes}
\end{figure}

Neutrino fluxes feature astrophysical uncertainties as well as potential contribution from flavor transitions induced by neutrino self-interactions \cite{Duan:2010bg}. 
During deleptonization phase of SNe where it is essentially only electron neutrinos that get produced, the effects from self-interactions are suppressed, particularly for the ``standard'' iron core SNe \cite{Duan:2007sh}; collective effects for such early neutrinos are non-existent \cite{Hannestad:2006nj}. The standard matter potential hence dominates and it induces the occurrence of two Mikheyev-Smirnov-Wolfenstein resonances \cite{Wolfenstein,Mikheev:1986gs,Mikheev:1986wj}, corresponding to the atmospheric and solar mass squared differences, which neutrinos encounter while travelling outwards\footnote{In contrast, matter effects from neutrino propagation in Earth are small and can be ignored \cite{Linzer:2019swe}.}. 
Provided the adiabatic evolution, the fluxes of neutrinos \cite{Lu:2016ipr,Dighe:1999bi} at distances much greater than the radius of the star read

\begin{align}
\Phi_{\nu_e}&= \Phi^0_{\nu_x}\,, \nonumber \\ 
\Phi_{\bar{\nu}_e}&= \Phi^0_{\bar{\nu}_e} \cos^2 \theta_{12} + \Phi^0_{\nu_x} \sin^2 \theta_{12} \,, \nonumber \\ 
\Phi_{\nu_x}&= \frac{1}{4}\, (2+\cos^2 \theta_{12}) \,\Phi^0_{\nu_x} +\frac{1}{4} \Phi^0_{\nu_e} +\frac{1}{4} \Phi^0_{\bar{\nu}_e} \sin^2 \theta_{12}\,,
\label{eq:fluxes-NO}
\end{align}
for NO and

\begin{align}
\Phi_{\nu_e}&= \Phi^0_{\nu_e} \sin^2 \theta_{12} + \Phi^0_{\nu_x} \cos^2 \theta_{12}\,, \nonumber \\ 
\Phi_{\bar{\nu}_e}&=  \Phi^0_{\nu_x}  \,, \nonumber \\ 
\Phi_{\nu_x}&= \frac{1}{4}\, (2+\sin^2 \theta_{12}) \,\Phi^0_{\nu_x} +\frac{1}{4} \Phi^0_{\bar{\nu}_e} +\frac{1}{4} \Phi^0_{\nu_e} \cos^2 \theta_{12}\,,
\label{eq:fluxes-IO}
\end{align}
 for IO where $\theta_{12}$ stands for the solar mixing angle and $\nu_x$ represents fluxes of muon and tau (anti)neutrinos that are practically identical in SNe. The fluxes in \cref{eq:fluxes-NO,eq:fluxes-IO} follow inverse-square law; in this paper we assume galactic SN at the distance of $d=10$ kpc from Earth. $\Phi^0$ represents fluxes at distances closer to the center of the star than the two resonance regions. For $\Phi^0$, we adopt parametrization from \cite{Keil:2002in} involving the so called pinching parameter through which the deviation of neutrino distribution function from the Maxwell-Boltzmann one is quantified, as well as the neutrino luminosity, $L$, ($\Phi^0 \propto L$) that is a parameter commonly included in the output of SN simulations. In \cref{fig:fluxes} we show neutrino luminosities for all neutrino flavors. In the left panel we focus on early times ($t<20$ ms, where $t=0$ is defined by the time of the core bounce) and it can be inferred that in this time window  luminosities are practically model independent. To illustrate that, we employed four SN simulations from \cite{Hudepohl2013,Mirizzi:2015eza}, see also 
\cite{Brdar:2018zds}. In contrast, in the right panel we show luminosities associated to the accretion and cooling phase; clearly, the theoretical uncertainties in this time window are much larger. The cutoff for SN3 and SN4 models arises because in those simulations star collapses into a black hole \cite{Beacom:2000qy}. For our numerical calculations, we use model denoted as SN2, obtained by simulating the core-collapse of a 27 $M_\odot$ progenitor star.
From the left panel it is obvious that electron neutrinos strongly dominate  up to roughly 10 ms after the core bounce when neutronization burst \cite{Kachelriess:2004ds} ends and we stress that it is precisely neutrinos produced in this time window that are crucial for the success of the method proposed in this work, as discussed below.

\begin{figure}
  \centering
 \begin{tabular}{cc}
    \includegraphics[width=0.98\textwidth]{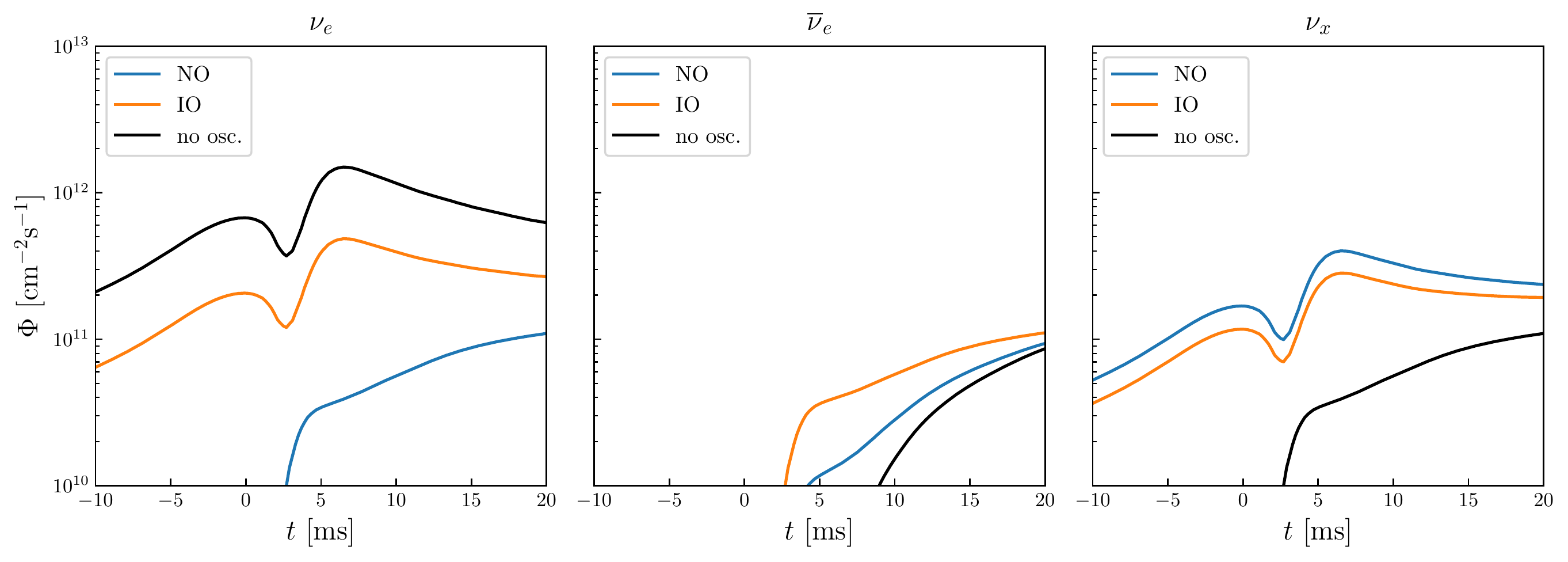} 
  \end{tabular}
  \caption{Comparison between SN neutrino flux in NO (blue) and IO (orange) as well as in the absence of neutrino oscillations (black).}
  \label{fig:fluxes2}
\end{figure}

 Following the transition of resonance regions, SN neutrinos that will be observed from this early emission will not be $\nu_e$ in case of NO (see again \cref{eq:fluxes-NO}). For IO, on the other hand, the difference between these early $\nu_e$ fluxes at smaller and larger distances (modulo inverse-square law dependence) with respect to resonance locations is only a factor of $\sin^2 \theta_{12} \approx 0.25$. We illustrate that in \cref{fig:fluxes2}. There, for each neutrino flavor (see particular panel) we show fluxes before (black) and after (orange for IO, blue for NO) encountering resonances. Focusing on the left panel it becomes clear that for a detector that will be particularly successful in detecting $\nu_e$, more of early produced neutrinos will be detected if the mass ordering is IO. In turn, the onset of SN events in such a detector should occur earlier in time for IO.

DUNE is such an experiment that will predominantly detect SN $\nu_e$ via the charged-current process with Ar (ArCC): $\nu_e+^{40}\text{Ar} \to e^- + ^{40}\text{K}^*$.  In contrast, other large neutrino detectors like JUNO and Super/Hyper-Kamiokande have greater capability of detecting $\overline{\nu}_e$ via the IBD process.
To investigate their capability for SN neutrino detection, we need to compute the event rates,  determined by
 \begin{align}
R(t)= N_{\text{target}}\,\int \Phi(E_\nu,t) \sigma(E_\nu) dE_\nu\,, 
 \label{eq:event_rates}
\end{align} 
where $N_{\text{target}}$ is the number of target particles in the detector and $\sigma(E_\nu)$ is the cross section for the detection process.
We employ this formula for all channels and all (anti-)neutrino species by properly taking fluxes, cross sections and number of target particles for different detectors. 
Apart from the two dominant detection channels (ArCC and IBD), we also consider all-flavor neutrino-electron elastic scattering events (eEs) for the aforementioned detectors as well as neutrino-proton elastic scattering events (pES) for JUNO. 
The cross sections for eES and pES are taken from Refs.~\cite{Beacom:2002hs,Giunti}.  The thresholds of electron recoil are set to 5, 0.2, and 5 MeV for DUNE, JUNO, and Super/Hyper-Kamiokande, respectively~\cite{Capozzi:2018dat, Li:2017dbg}. As for pES, using the quenched energy deposit indicated in Fig.~5 in Ref.~\cite{Beacom:2002hs}, we set the threshold of proton recoil at 1.2 MeV for JUNO.
The fiducial masses of DUNE, JUNO, and Super/Hyper-Kamiokande are set to 40, 20, and 32/374 kt respectively. For DUNE, such mass will be obtained by successively adding individual 10 kt modules at the far detector. In addition, in all our calculations we include background events \cite{Brdar:2018zds}. 
With the above setup, we compute the expected number of galactic SN events for the aforementioned experiments; these are shown in \cref{tab:events}.

\begin{table}
  \begin{tabular}{c|c|c|c|c}
  \hline 
   & $N$ ($t<20$ ms, NO)  & $N$ ($t<20$ ms, IO) & $N$ (total, NO) & $N$ (total, IO)  \tabularnewline
   \hline
  DUNE-ArCC & 11.3 & 50.9 & 3285 & 3097 \tabularnewline
  DUNE-eES & 2.99 & 6.48 & 311 & 314 \tabularnewline
  JUNO-IBD & 14.2 & 27.2 & 6297 & 6194 \tabularnewline
  JUNO-eES & 4.11 & 8.50 & 362 & 369 \tabularnewline
  JUNO-pES & 18.8 & 19.2 & 3670 & 3798 \tabularnewline 
  SuperK-IBD & 17.6 & 33.8 & 7830 & 7701 \tabularnewline
  SuperK-eES & 2.95 & 6.39 & 307 & 310 \tabularnewline
  HyperK-IBD & 206 & 395 & 91517 & 90011 \tabularnewline
  HyperK-eES & 34.5 & 74.7 & 3588 & 3628 \tabularnewline
  \hline
  \end{tabular}
  \caption{Number of events for indicated detectors and channels given in two time windows. In second (NO) and third (IO) column we focus on $t<20$ ms  while in the last two columns we show the total number of events associated to a SN explosion.}
    \label{tab:events}
  \end{table}

The main idea in this paper is to use both DUNE and JUNO or Super/Hyper-Kamiokande simultaneously in order to discover the mass ordering. Given the event rates in \cref{tab:events}, for IO there are clearly many more ArCC events
in $t<20$ ms window; this implies that for IO the first event recorded in the detector will occur earlier. For IBD channel, we do not see a dramatic difference between event counts across NO and IO for $t<20$ ms. Therefore, the time difference between the onset of neutrino events at DUNE and at a given IBD detector will be larger for IO (after appropriately subtracting time of propagation in Earth for two detectors at different locations) and this is the reason how one can probe mass ordering by focusing only on timing of neutrino events. In \cref{sec:results} we will introduce statistical method for determination of the time of $n$-th neutrino event in the detector which will allow us to assess how many neutrino events are required in order to make interesting statistical statements on the mass ordering. Let us close this section by stating that in contrast to comparing ArCC and IBD events at different detectors one may be tempted to also compare interaction times of first few eES and ArCC at DUNE or eES and IBD at JUNO or Super/Hyper-Kamiokande, focusing hence on a single detector and two channels. This is possible since events from charged current scattering on nuclei and those from neutral current scattering on electrons can easily be distinguished from one another. While we will also show projections for such strategy in \cref{sec:results}, we will eventually conclude that the most promising situation for determination of the ordering is still the case where two different detectors are involved.

\section{Statistical Methods and Results}
\label{sec:results}

\begin{figure}
  \centering
 \begin{tabular}{cc}
    \includegraphics[width=0.95\textwidth]{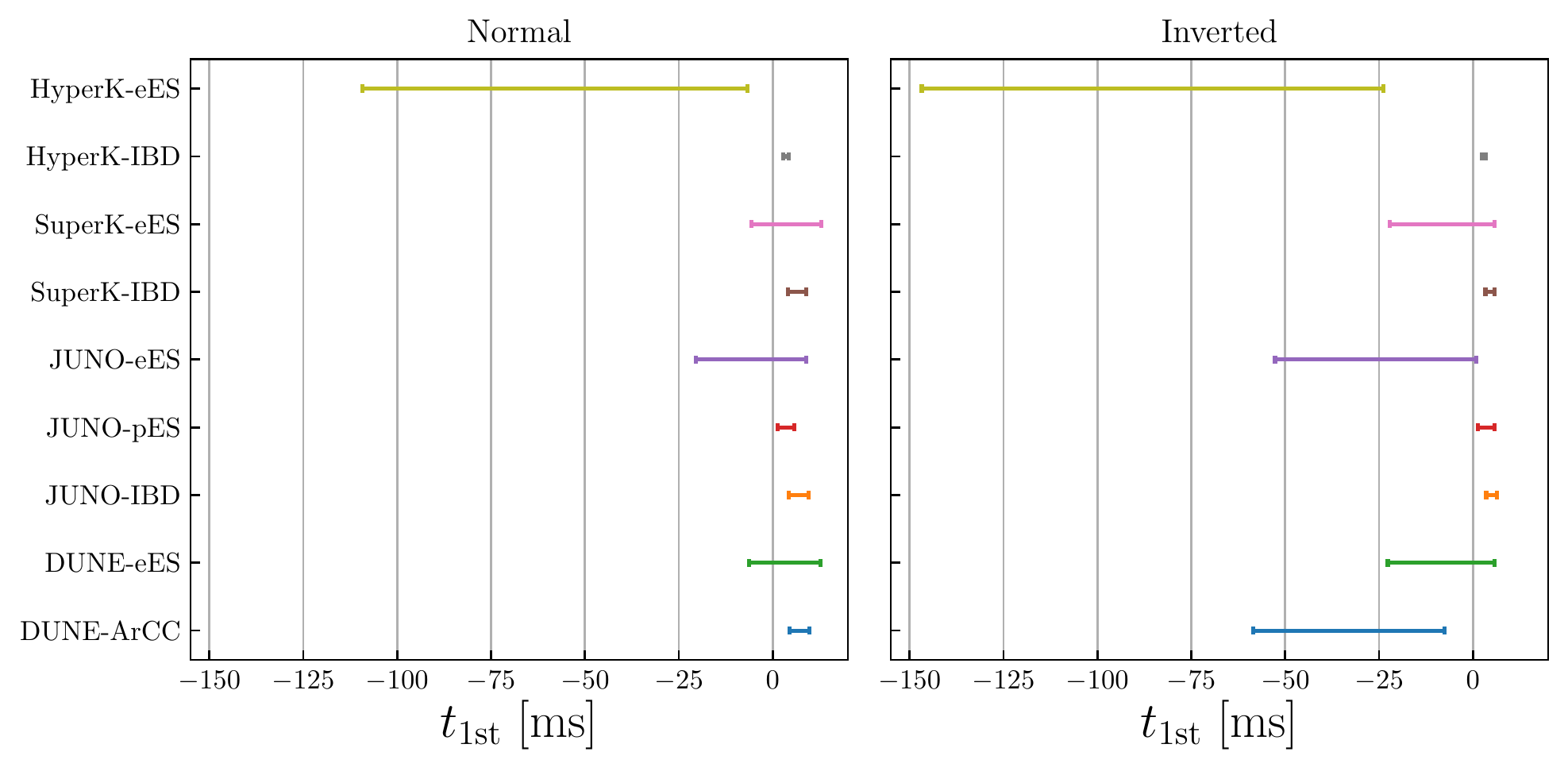} \\
    \includegraphics[width=0.95\textwidth]{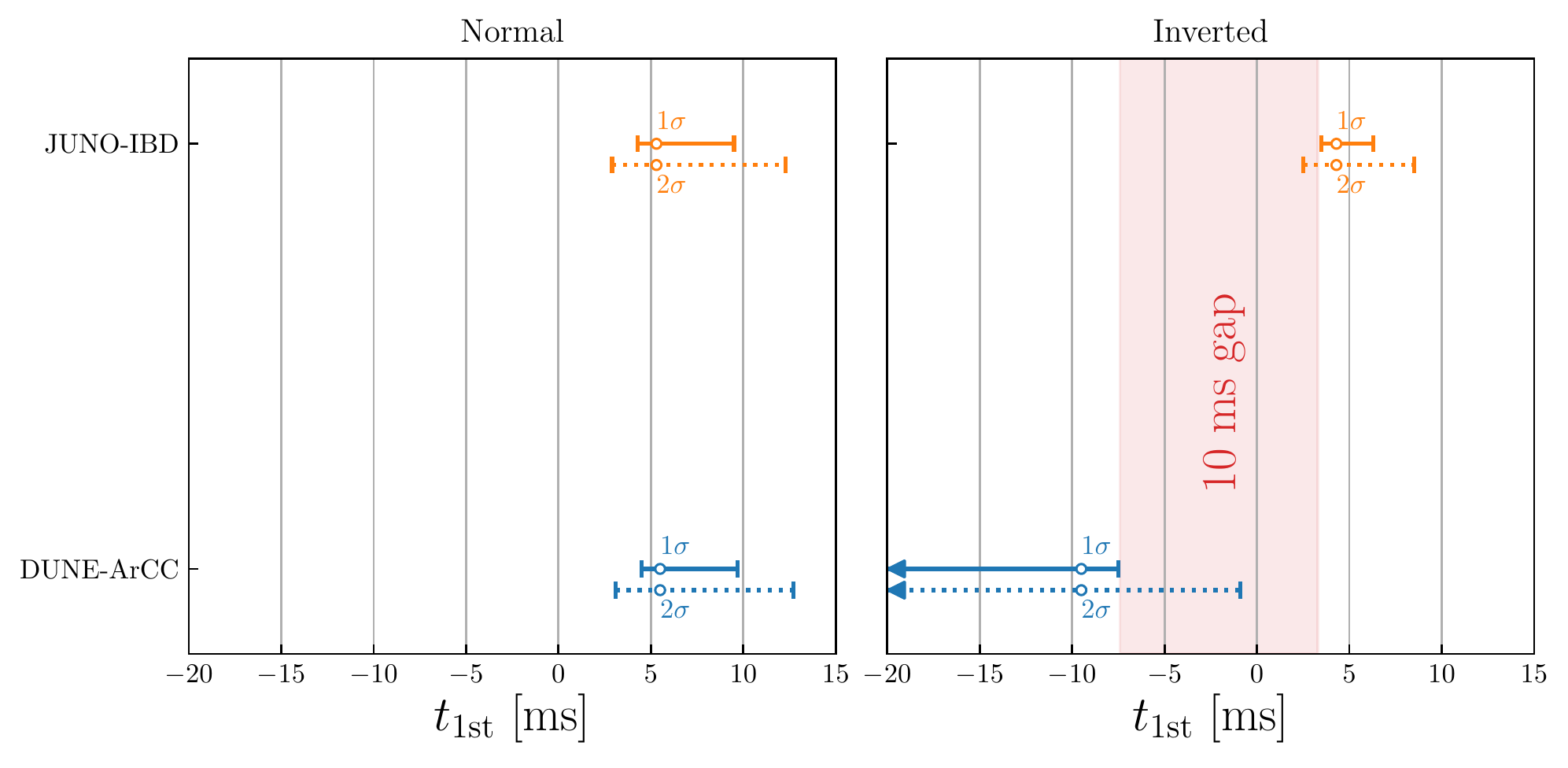}
  \end{tabular}
  \caption{Expected time window for the first event detection. The upper panels show 1$\sigma$ uncertainty bars for the time  of the first event detection ($t_{\rm 1st}$) at DUNE, JUNO and  Super/Hyper-Kamiokande via multiple detection channels including ArCC, IBD, eES, and pES. Left (right) panel is for NO (IO). Lower panels are dedicated to the comparison between JUNO-IBD and DUNE-ArCC to illustrate that the time separation between the respective first events is typically $\gtrsim 10$ ms in the IO, which would be a statistically significant signal for mass ordering. In lower panels we also show the maximal probability points (marked by ``$\circ$'') and  2$\sigma$ uncertainty bars.
  }
  \label{fig:bar}
\end{figure}

\noindent
Let us consider the first event in a detector and denote the time
of the occurrence of such an event by $t_{{\rm 1st}}$. Given the
expected event rate, $R(t)$, we would like to know  the statistical
expectation and fluctuation of $t_{{\rm 1st}}$. Obviously, the first
event is unlikely to appear in any $R(t)$-suppressed period. It is
also unlikely to appear when the integrated event rate is too large
($\int^{t}R(t')dt'\gg1$), because it implies that a large number
of events should have already occurred. In Appendix~\ref{sec:statistics} we have
computed the probability density function (p.d.f) of $t_{{\rm 1st}}$ which reads
\begin{equation}
p_{{\rm 1st}}(t_{{\rm 1st}})=R(t_{{\rm 1st}})\exp\left[-\int_{-\infty}^{t_{{\rm 1st}}}R(t)dt\right].\label{eq:pdf}
\end{equation}
It is indeed suppressed when either the event rate is small or
the integrated event rate is large. 

Using Eq.~\eqref{eq:pdf}, we can compute the expected interval of
$t_{{\rm 1st}}$ for specified CL. Denote
the probability of $t_{{\rm 1st}}$ occurring in $[t_{-},\ t_{+}]$
by $1-\alpha$, where $\alpha=0.3173$ ($0.0455$) for 1$\sigma$
(2$\sigma$) CL. Then, the interval $[t_{-},\ t_{+}]$ is determined
by 
\begin{equation}
\int_{-\infty}^{t_{-}}p_{{\rm 1st}}(t)dt=\int_{t_{+}}^{\infty}p_{{\rm 1st}}(t)dt=\frac{\alpha}{2}.\label{eq:alpha_to_t}
\end{equation}

In the upper panels of Fig.~\ref{fig:bar}, we show 1$\sigma$ intervals of $t_{\rm 1st}$ for all important detection channels in DUNE, JUNO and Super-/Hyper-Kamiokande. One can see that the intervals for the IBD channel, through which only $\overline{\nu}_e$ are detected, are comparatively late and short in both NO and IO cases, while the  $\nu_e$-dedicated channel ArCC shows a significantly earlier interval in the IO. For pES, since the cross sections of neutrinos and anti-neutrinos are almost the same\footnote{
  The difference between $\nu+p$ and $\overline{\nu}+p$ cross sections is suppressed by $E_{\nu}/m_p$~\cite{Beacom:2002hs}. 
} and independent of flavors, the 1$\sigma$ intervals in the NO and IO are nearly identical. For eES, neutrinos and anti-neutrinos of all flavors are detected and, due to the differences of cross sections ($\sigma_{\nu_e+e}> \sigma_{\overline{\nu}_e+e} > \sigma_{\nu_{x}+e}> \sigma_{\overline{\nu}_x+e} $), 1$\sigma$ intervals are sensitive to mass ordering. 

In the lower panels of Fig.~\ref{fig:bar}, we concentrate on the comparison between DUNE-ArCC and JUNO-IBD measurements to further illustrate our method. As shown in the lower left panel, the 1$\sigma$ intervals of $t_{\rm 1st}$ are very similar for NO: $t_{\rm 1st}\in [4.5, 9.7]$ ms for JUNO-IBD, and $[4.3, 9.5]$ ms for DUNE-ArCC. This implies that the time difference  between the first events at JUNO-IBD and DUNE-ArCC in the NO case is likely to be less than $\sim 5$ ms at 1$\sigma$ CL.  In the IO case, as shown in the lower right panel, there is a significant gap between the 1$\sigma$ intervals: $t_{\rm 1st} \in [3.5, 6.3]$ ms for JUNO-IBD and $[-58.5, -7.5]$ ms for DUNE-ArCC, which implies that the time difference should be greater than $\sim 10$ ms at 1$\sigma$ CL. Note that, as a consequence of rapidly increasing event rates,  the p.d.f of $t_{\rm 1st}$ is non-Gaussian and the maximal probability points are not in the middle of these intervals.  

In order to obtain the statistical significance for discriminating between NO and IO reflected in Fig.~\ref{fig:bar}, we shall inspect the statistics of the time difference of the first event ($\Delta t\equiv t_{\rm 1st}^{a}-t_{\rm 1st}^{b}$) between two experiments labelled as $a$ and $b$. Given the p.d.f of $t_{\rm 1st}^{a}$ and $t_{\rm 1st}^{b}$, which are denoted by  $p_a$ and $p_b$ respectively, the p.d.f of  $\Delta t$ is determined by (see Appendix~\ref{sec:statistics})
\begin{equation}
  p_{\Delta}(\Delta t)=\int p_{a}\left(\frac{t+\Delta t}{2}\right)p_{b}\left(\frac{t-\Delta t}{2}\right)\frac{dt}{2}\thinspace.
  \label{eq:delta_t}
\end{equation}
	Using Eq.~\eqref{eq:delta_t}, we find that NO and IO scenarios are statistically discernible at 1.8$\sigma$ (2.4$\sigma$) CL by employing only the first SN events at JUNO-IBD (HyperK-IBD) and DUNE-ArCC. 
 In \cref{sec:pr} we further show that when the neutrino fluxes from the Garching group are replaced by those calculated by those from the Princeton group, the statistical significance changes to  2.6$\sigma$ (2.7$\sigma$).

\begin{figure}
  \centering
 \begin{tabular}{cc}
  \includegraphics[width=0.88\textwidth]{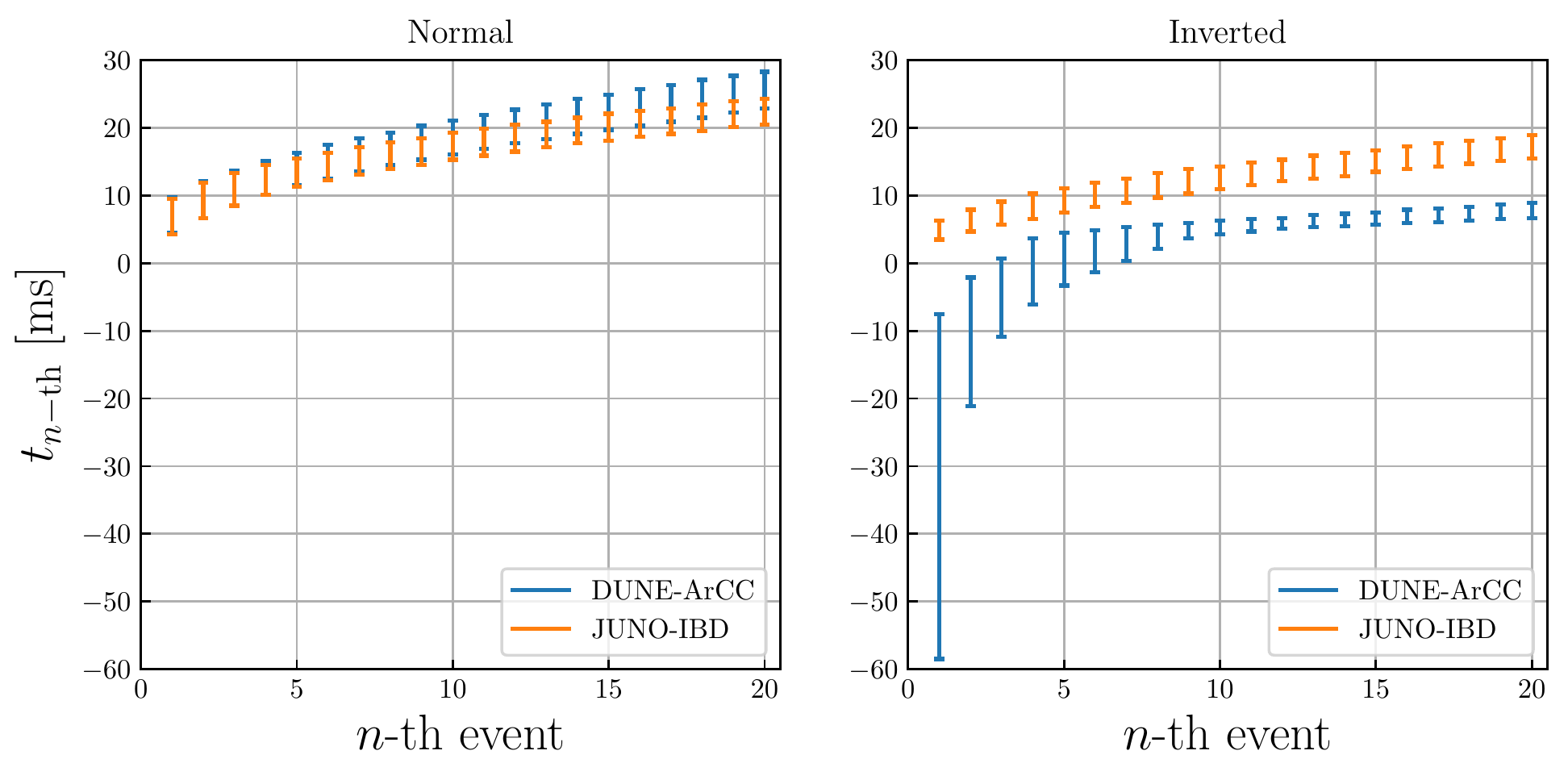} 
  \end{tabular}
  \caption{
   The expected time of first 20  events at DUNE and JUNO in the ArCC and the IBD detection channels, respectively. Left (right) panel is for NO (IO).
  }
  \label{fig:many_bar}
\end{figure}

\begin{figure}
  \centering
 \begin{tabular}{cc}
    \includegraphics[width=0.88\textwidth]{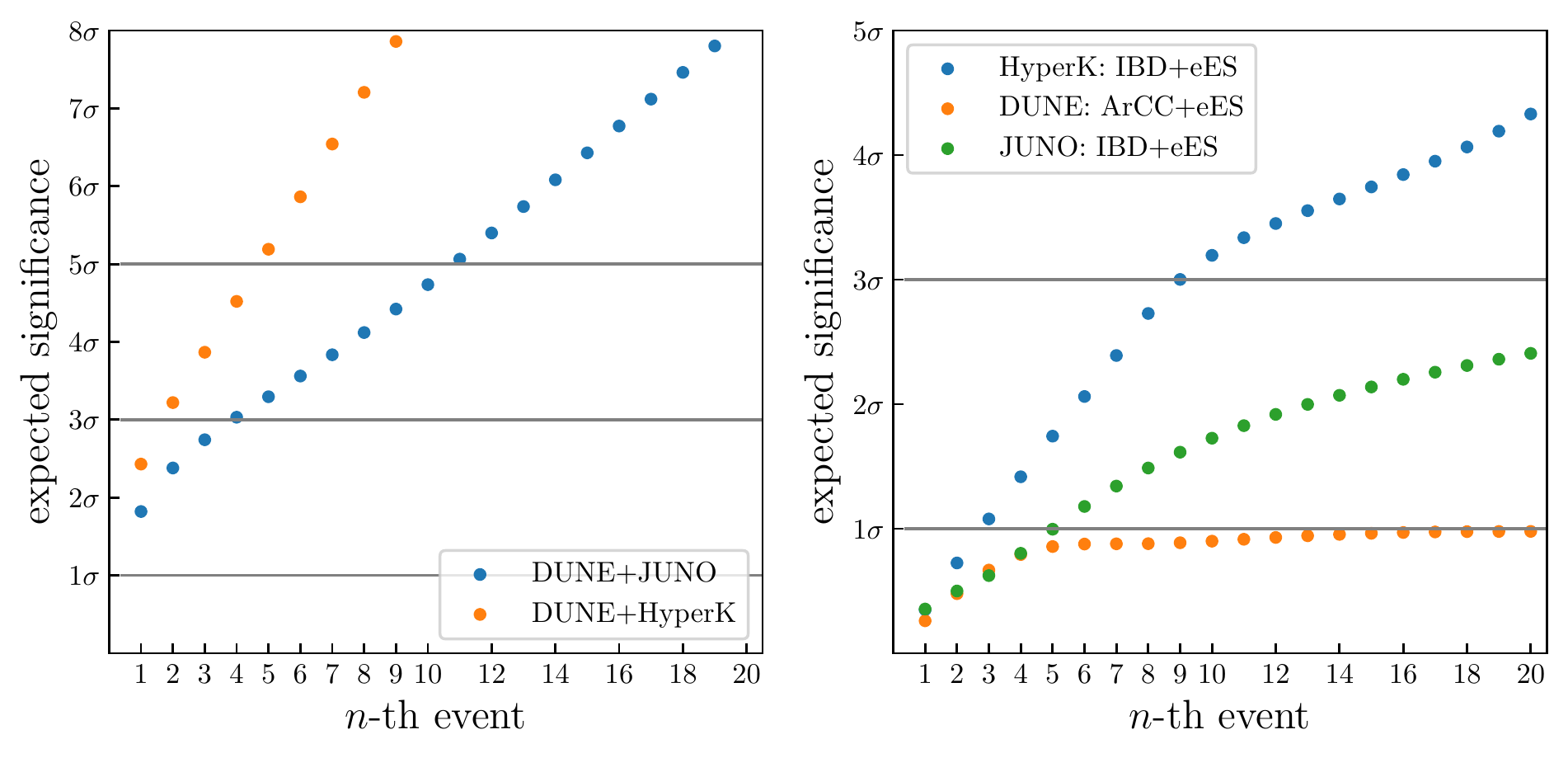} 
  \end{tabular}
  \caption{Expected significance for neutrino mass ordering as a function of recorded number of events for which we only require the timing information. In the left panel, ArCC in DUNE and IBD in JUNO or Hyper-Kamiokande is considered. Knowing only the timing of the first event in these two channels, a $2\sigma$ statement can be made. For a discovery with DUNE+Hyper-Kamiokande (DUNE+JUNO) 5 (11) events suffice. In the right panel we compare charged current channel and neutrino-electron elastic scattering channel for all detectors separately and find that only at Hyper-Kamiokande $>3\sigma$ statement can be made. This implies that the interplay of two different detectors by comparing respective dominant interaction channels (ArCC and IBD) is most successful for probing the mass ordering.}
  \label{fig:n-sigma}
\end{figure}

Going beyond the first-event analysis, one can significantly improve the results by  including more subsequent events.
In Fig.~\ref{fig:many_bar}, we show the 1$\sigma$ intervals of $t_{n\text{-th}}$, which is defined as the time of the $n$-th event. For the first 20 events, DUNE-ArCC events will appear significantly earlier than JUNO-IBD ones in the IO case; in contrast, as discussed above, the timing in these two channels for NO is similar. 
Applying Eq.~\eqref{eq:delta_t} to obtain the p.d.f of $\Delta t$ for the $n$-th events, we can obtain the statistical significance of using the $n$-th events to discriminate between NO and IO, similar to the above analysis for the first events.  
Assuming that the time distributions of these events are statistically independent\footnote{
	We expect  however that among these time-ordered events there should be correlations, which are not straightforward to take into account in the analysis. Our negligence of the correlations might lead to  somewhat overoptimistic results.}, 
we simply add up their contributions (in terms of the corresponding log-likelihood values) to obtain the cumulative statistical significance. 
The results are presented in Fig.~\ref{fig:n-sigma}. 

We find that with the first 4 (11) events from  DUNE-ArCC and JUNO-IBD, NO and IO can be distinguished at 3$\sigma$ (5$\sigma$) CL --- see the left panel in Fig.~\ref{fig:n-sigma}. The combination of DUNE-ArCC with SuperK-IBD leads to very similar results. If DUNE-ArCC is combined with HyperK-IBD, 5$\sigma$ CL can be reached with the first 5 events.  For the Princeton fluxes, the number of events required for  5$\sigma$ CL will be changed to 20 and 7 for DUNE-JUNO and DUNE-Hyper-Kamiokande combinations, respectively (see \cref{sec:pr}).

In the above analyses, we assume that the time differences from geographical locations of  detectors have been subtracted, which is feasible if the direction of the SN neutrino flux is known (e.g. if the SN has been optically observed or if its astronomical coordinates have been reconstructed from eES \cite{Super-Kamiokande:2016kji}). Even without knowing the direction, one can still compare two different channels in the same detector. In the right panel of Fig.~\ref{fig:n-sigma}, we show the results of combining IBD or ArCC with eES in a single detector. Since in each combination either $\nu_e$ or $\overline{\nu}_e$ detection is limited by low statistics of (anti)neutrino-electron scattering events, the capability of determining the mass ordering is weaker when compared to the inter-detector combinations. For IBD+eES in JUNO, one can get $\sim 1 \sigma$ significance with the first 5 events. 
For ArCC+eES in DUNE, the significance of the first 4 events is similar but with more subsequent events it increases slower than that for IBD+eES in JUNO. We note that this depends on the SN models simulated by different group (c.f. Fig.~\ref{fig:n-sigma-pr}). 
 For IBD+eES in Hyper-Kamiokande, due to its much higher statistics,  $3 \sigma$ significance can be reached with the first 9 events. This conclusion remains robust across considered SN models.

\section{Interplay with SN Triangulation }
\label{sec:triangulation}
\noindent
The idea of SN triangulation, i.e.~locating the position of the SN in the sky by using 
the time differences of onset of neutrino events across several detectors, was first discussed in \cite{Beacom:1998fj} where the scepticism about this method was presented. The idea was, however, revived in \cite{Brdar:2018zds} where it was shown, using fluxes from modern SN simulations, that the statistical uncertainties at present and near future detectors are such that triangulation becomes feasible. It was found that the galactic SN can be located with $\lesssim 5^\circ$ in both declination and right ascension coordinates. Later, in Ref. \cite{Linzer:2019swe} (see also \cite{Muhlbeier:2013gwa,Coleiro:2020vyj,Sarfati:2021vym} and 
related work on supernova timing \cite{GFtiming,ICtiming,Hansen:2019giq}), the success of triangulation was further verified and now this methods is one of the prominent goals of SNEWS \cite{SNEWS:2020tbu}.

Here we wish to briefly discuss the interplay between triangulation and mass ordering determination. As discussed in \cref{sec:results}, the method that guarantees 5$\sigma$ CL result for the mass ordering involves the usage of event time differences between different detectors such as DUNE and JUNO, or DUNE and Hyper-Kamiokande (see the left panel in \cref{fig:n-sigma}).
The actual measurement of such time differences should be corrected by the propagation time difference arising from different locations of detectors with respect to the SN. 
It is crucial to account for the propagation time difference because it can be as large as $\mathcal{O}(10)$ ms. For DUNE and JUNO (Hyper-Kamiokande), the distance between the two detectors is 10.2 (8.4)$\times 10^3$ km, respectively. If the SN direction is in alignment with the connection of the two detectors, then the geographical contribution to the time difference is 34 (28) ms. Since the triangulation method makes use of these time differences, one can envisage that triangulation and mass ordering determination would be considerably entangled. 

One way to disentangle the two would be to make use of the angular dependence of interaction channels such as eES which can determine the direction up to a few degrees\footnote{Another promising and potentially very precise method for determining the position of galactic supernova in the sky is by detecting photons via optical telescopes. However, there are obstacles such as the limited field of view of such telescopes \cite{SNEWS:2020tbu}; note also that in scenarios where the core of a SN collapses into a black hole there is no optical signal.} \cite{Super-Kamiokande:2016kji}. Taking, e.g., $5^{\circ}$ for the uncertainty of the direction, the geographical time difference can be determined with an uncertainty of $\sim 2.6$ ms, which would be a subdominant uncertainty when compared to the error bars in \cref{fig:many_bar}. Hence, our method in \cref{sec:results} for determination of the mass ordering is rather feasible. 
	A potential concern in using the eES events to disentangle mass ordering and triangulation is that the reconstruction of the direction of eES events may take considerably longer time than triangulation. By the time the eES direction is reconstructed, an early analysis of triangulation based on time characteristics independent of mass ordering (e.g. the position of the neutronization burst peak, instead of first events) may have already been completed. In this case, the early triangulation result can be used to improve both the determination of mass ordering as well as to perform more elaborated calculations of triangulation which are mass-ordering dependent.
	
In particular, once the ordering is known, the triangulation method proposed in \cite{Linzer:2019swe} can be performed. Looking at \cref{fig:many_bar}, when considering the first few events, 
the statistical uncertainty of the $n$-th event time at DUNE is quite large for IO. The triangulation method advocated in \cite{Linzer:2019swe} is based on the comparison of the time of first neutrino events in several detectors. This turns out to be inefficient for IO; in that case it is much more advantageous to skip the first event and  focus on later events (for instance the 10th event has a much smaller uncertainty and is still arising from the period of neutronization). On the other hand, if various mass ordering measurements converge to NO, then the first-event method should be robust.

\section{Summary and Conclusions}
\label{sec:conclusion}
\noindent
In this work we have proposed a novel method for determination of neutrino mass ordering using SN neutrinos. Compared to previous studies based on high-statistics measurement of SN neutrino fluxes, our method focuses exclusively on the time information on the first couple of events during the period of infall and neutronization burst. The neutrino fluxes in this period are known to be the least dependent on SN models (see Fig.~\ref{fig:L-t}). 

The crucial point for the success of this approach is the synergy among several upcoming
large detectors including DUNE, JUNO and Hyper-Kamiokande. 
In the case of IO, a large time gap ($\sim 10$ ms) between the onset of SN events in DUNE and JUNO/Hyper-Kamiokande is expected (see Fig.~\ref{fig:bar}) while for NO all aforementioned detectors should start recording events almost simultaneously, due to the suppressed flux of electron neutrinos produced in the infall phase of SN. 
 We found that recording only the first $10\sim 20$ events suffices for conclusive ($5\sigma$) determination  of the mass ordering while the time difference between the very first event at DUNE and IBD detectors is already enough for a $2\sigma$ statement, as shown in Fig.~\ref{fig:n-sigma}.
 
Our method can be readily incorporated in SNEWS which will have immediate access to the first few events from aforementioned detectors when the next galactic SN occurs. Provided the SN position is known from e.g.~optical observations and/or from the  observed angular dependence of neutrino interaction channels, the proposed method will allow SNEWS to determine mass ordering with these early events. We have also discussed the interplay between mass ordering and determination of the SN position in the sky via triangulation and argued that the triangulation method can be successfully performed following determination of the mass ordering.

\section*{Acknowledgements} 
\noindent
We benefited greatly from discussions with Alexander Friedland, Shunsaku Horiuchi, Shirley Li and Shun Zhou. We are also grateful to Alec Habig and Kate Scholberg for reading the final version of the manuscript. Fermilab is operated by the Fermi Research Alliance, LLC under contract No. DE-AC02-07CH11359 with the United States Department of Energy. This work was performed in part during KITP program 
``Neutrinos as a Portal to New Physics and Astrophysics''.  X.J.X is supported in part by the National Natural Science Foundation of China under Grant No. 12141501.

\appendix

\section{The statistical uncertainties of early events\label{sec:statistics}}

In this appendix, we derive the formula used to study the statistical
fluctuation of the time of the first event, $t_{{\rm 1st}}$, and
then generalize it to the second, third and other early events that
could be of importance to our analysis.

Given an event rate curve $R(t)$ like the one computed in Eq.~\eqref{eq:event_rates},
the expected number of events occurring before time $t$  reads 
\begin{equation}
\mu(t)=\int_{-\infty}^{t}R(t')dt'\thinspace,\label{eq:mu}
\end{equation}
where the lower bound does not necessarily need to correspond to $-\infty$; it can be any time below which $R(t')$ vanishes or is negligible.   In an
actual observation, one can only detect an integer number of events
and the probability of detecting $n$ events is governed by the Poisson
distribution
\begin{equation}
P_{\mu}(n)=\frac{\mu^{n}}{n!}e^{-\mu}\thinspace.\label{eq:poiss}
\end{equation}
The probability of the first event occurring within $[t,\ t+dt]$
should be the probability of \emph{no} events occurring before $t$
multiplied by the probability of \emph{one} event in $[t,\ t+dt]$ interval
\begin{equation}
P\left(t_{{\rm 1st}}\in[t,\ t+dt]\right)=P_{\mu}(0)P_{d\mu}(1)\thinspace,\label{eq:p1st}
\end{equation}
where   $d\mu=R(t)dt$ according to Eq.~\eqref{eq:mu}.

Substituting Eq.~\eqref{eq:poiss} into Eq.~\eqref{eq:p1st} we obtain
the probability density function of $t_{{\rm 1st}}$ 
\begin{equation}
p_{{\rm 1st}}(t_{{\rm 1st}})=R(t_{{\rm 1st}})\exp\left[-\int_{-\infty}^{t_{{\rm 1st}}}R(t)dt\right].\label{eq:}
\end{equation}
Note that 
\begin{equation}
\int_{-\infty}^{t}p_{{\rm 1st}}(t_{{\rm 1st}})dt_{{\rm 1st}}=1-e^{-\mu}\, ,\label{eq:-2}
\end{equation}
which is expected since the left-hand side corresponds to the probability
of the first event occurring before $t$ while $e^{-\mu}=P_{\mu}(0)$
on the right-hand side is the probability of no events occurring before
$t$. The two probabilities should be complementary to each other.

Next, let us generalize Eq.~\eqref{eq:} to the case of the second
event, for which the time is denoted by  $t_{{\rm 2nd}}$. The probability
of the second event occurring within $[t,\ t+dt]$ should be the probability
that one (and only one) event has occurred before $t$ multiplied
by the probability of one event occurring within $[t,\ t+dt]$
\begin{equation}
P\left(t_{{\rm 2nd}}\in[t,\ t+dt]\right)=P_{\mu}(1)P_{d\mu}(1)\thinspace.\label{eq:p2nd}
\end{equation}
This gives
\begin{equation}
p_{{\rm 2nd}}(t_{{\rm 2nd}})=\mu(t_{{\rm 2nd}})\exp\left[-\mu(t_{{\rm 2nd}})\right]R(t_{{\rm 2nd}})\thinspace.\label{eq:-1}
\end{equation}

Further generalizations to the $n$-th event are straightforward.
The probability density function of $t_{n\text{-th}}$ reads
\begin{equation}
p_{n\text{-th}}(t_{n\text{-th}})=\frac{\mu(t_{n\text{-th}})^{n-1}}{(n-1)!}\exp\left[-\mu(t_{n\text{-th}})\right]R(t_{n\text{-th}})\thinspace.\label{eq:-3}
\end{equation}
Similar to Eq.~\eqref{eq:-2}, the integral of $p_{n\text{-th}}(t_{n\text{-th}})$
can be calculated analytically
\begin{equation}
\int_{-\infty}^{t}p(t_{n\text{-th}})dt_{n\text{-th}}=1-\frac{\Gamma(n,\mu)}{(n-1)!}\thinspace,\label{eq:-4}
\end{equation}
where $\Gamma(n,\mu)$ is the incomplete gamma function. 

For two independent events (e.g.~first events occurring at two
detectors), given their respective probability density functions $p_{a}(t_{a})$
and $p_{b}(t_{b})$, the probability density function of $t_{-}=t_{a}-t_{b}$
can be obtained by the transformation of random variables: $(t_{a},\ t_{b})\rightarrow(t_{-},\ t_{+})\equiv(t_{a}-t_{b},\ t_{a}+t_{b})$.
Including the Jacobian in this transformation ($dt_{-}dt_{+}=2dt_{a}dt_{b}$),
the probability density function of $t_{-}$ reads
\begin{equation}
p_{-}(t_{-})=\int p_{a}(t_{a})p_{b}(t_{b})\frac{dt_{+}}{2}=\int p_{a}\left(\frac{t_{+}+t_{-}}{2}\right)p_{b}\left(\frac{t_{+}-t_{-}}{2}\right)\frac{dt_{+}}{2}\thinspace.\label{eq:-5}
\end{equation}

\section{Results using SN neutrino spectra from a different group \label{sec:pr}}

\begin{figure}
	\centering
	\begin{tabular}{cc}
		\includegraphics[width=0.98\textwidth]{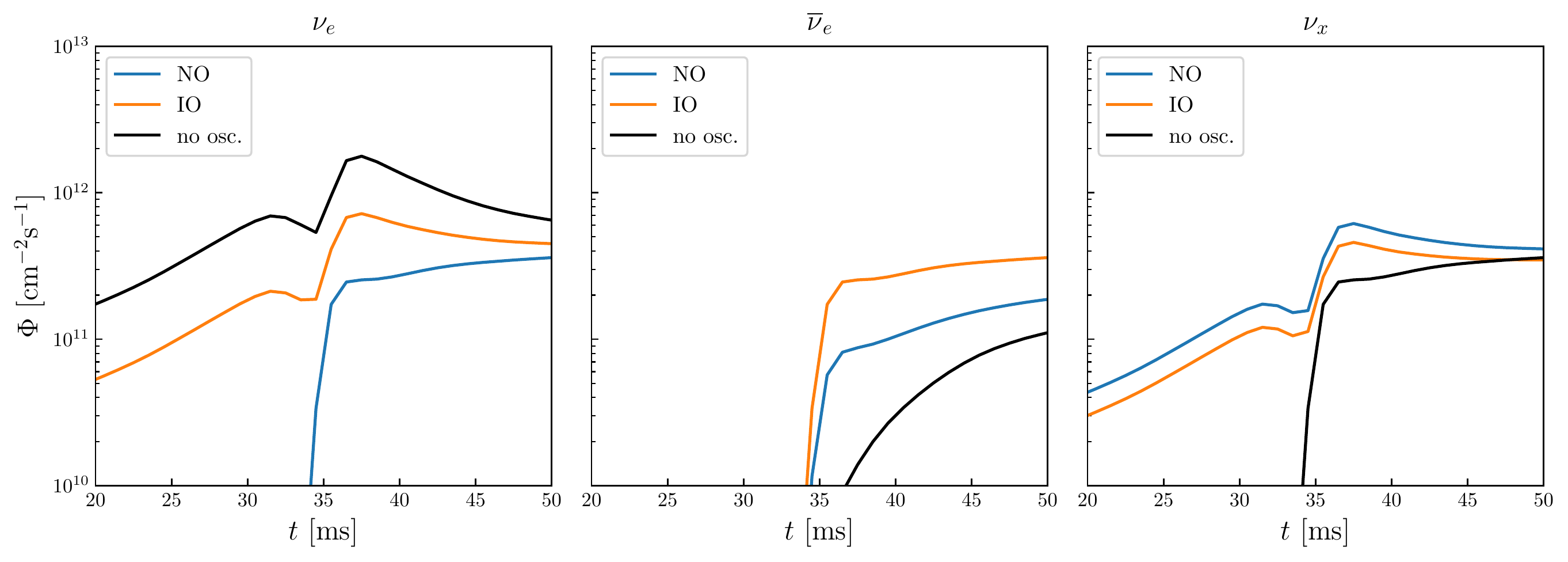} 
	\end{tabular}
	\caption{Similar to Fig.~\ref{fig:fluxes2} except for a SN model simulated by a different group~\cite{Burrows:2020qrp}.
	\label{fig:fluxes2-pr}}
	
%
%
	\centering
	\begin{tabular}{cc}
		\includegraphics[width=0.95\textwidth]{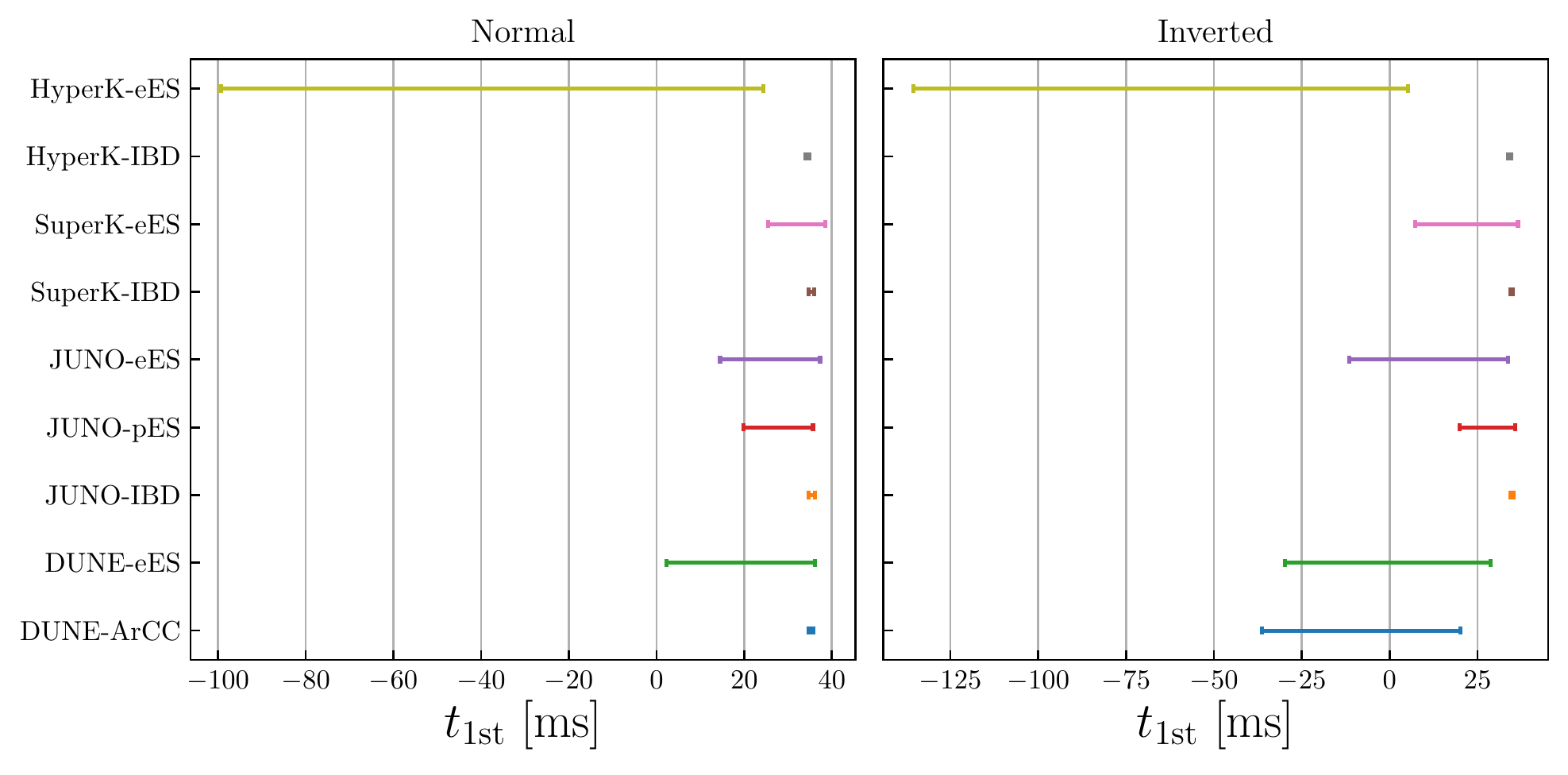}
	\end{tabular}
	\caption{Similar to the upper panels in Fig.~\ref{fig:bar} except for a SN model simulated by a different group~\cite{Burrows:2020qrp}.
	\label{fig:bar-pr}}
	
%
%
	\centering
	\begin{tabular}{cc}
		\includegraphics[width=0.88\textwidth]{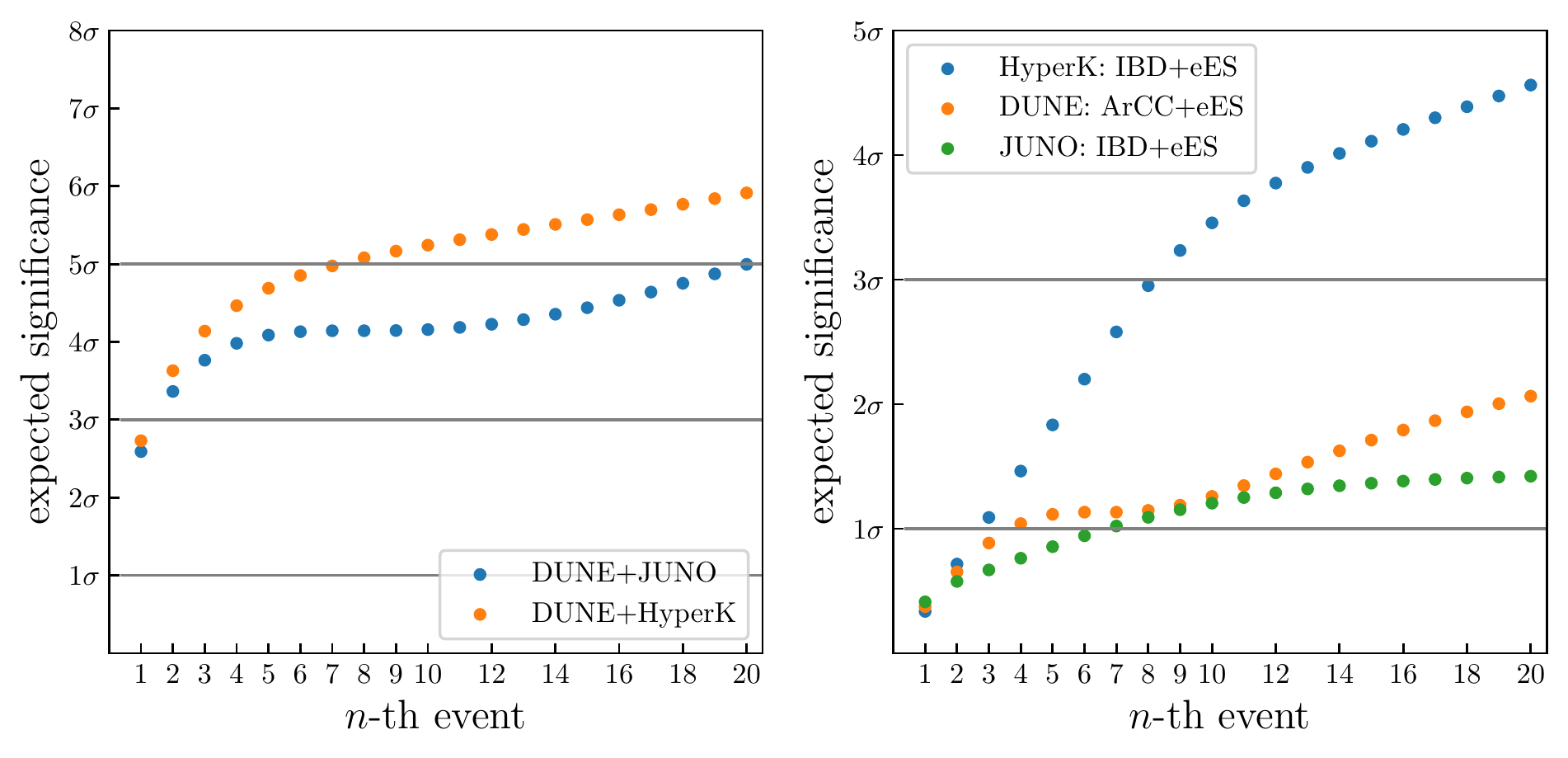} 
	\end{tabular}
	\caption{Similar to the upper panels in Fig.~\ref{fig:n-sigma} except for a SN model simulated by a different group~\cite{Burrows:2020qrp}.
	\label{fig:n-sigma-pr}}
	
\end{figure}

The results presented in the main text are obtained using the neutrino spectra from the Garching group~\cite{Hudepohl2013,Mirizzi:2015eza}.
Even though the SN simulations for different masses of progenitor stars performed by the same group yield very similar neutrino spectra, we have noticed that different groups find different spectra for the same progenitor mass. In order to illustrate how this might affect our conclusions, we have taken the neutrino spectra from the latest simulation by 
the Princeton group~\cite{Adam1,Adam2,Adam3, Burrows:2020qrp}. The data are available at the website\footnote{See \url{https://www.astro.princeton.edu/~burrows/nu-emissions.2d/data/}}, from where we take the spectra corresponding to 26 $M_\odot$ progenitor (notice that this is very close in mass to a 27 $M_\odot$ progenitor, simulation of which we use in this work in the context of Garching group neutrino fluxes). 
Let us state that we do not use the parametrization involving the pinching parameter because energy distributions are  already provided in the data set as 12-bin arrays for each $t$ point on the $t$ grid. Linear interpolation has been used in energy integrals and in calculations involving finer $t$ bins.
Using these fluxes, we simply repeat the analysis reported in the main text for Garching SN simulations. The results are presented in Figs.~\ref{fig:fluxes2-pr} to \ref{fig:n-sigma-pr}.

In Fig.~\ref{fig:fluxes2-pr}, we show the neutrino fluxes at Earth with and without flavor transition taken into account. Compared to Fig.~\ref{fig:fluxes2}, one can see that the NO curve in the leftmost panel is significantly higher than the corresponding curve from the Garching group. This feature generally reduces the sensitivity of our method because the high $\nu_e$ flux in the NO case causes $\nu_e$ events to appear at earlier time; however, notice from the last row in \cref{fig:bar-pr} that for ArCC events in DUNE there is still a sufficient time separation between occurrence of first event when comparing two different orderings. 

Quantitatively, the expected significance as a function of event number for this 
class of models is shown in \cref{fig:n-sigma-pr}. When comparing the left panel to the one in \cref{fig:n-sigma} one can infer that, when considering only first few events, the significance for 
Princeton
models is higher. This is because these models have much higher $\bar{\nu}_e$ fluxes when compared to Garching ones and that shrinks the error bar for the time of detection via IBD, leading to the increase in the significance for first few events. Note that subsequent events make smaller contributions to the statistical significance because, in the 
Princeton models, the fluxes of $\nu_e$ and $\bar{\nu}_e$ eventually reach the same order of magnitude. This happens shortly after $\bar{\nu}_e$ get significantly produced. Eventually, as seen from the figure, we find that by using DUNE+JUNO (DUNE+Hyper-Kamiokande) combinations 20 (7) events suffice for the discovery of the ordering and this is only slightly larger than our previous result using Garching fluxes (see again left panel \cref{fig:n-sigma}). In the right panel of \cref{fig:n-sigma-pr} we also show the expected significance when combining two detection channels for the same detector and find
very similar results to those reported in the right panel of  \cref{fig:n-sigma}. Ultimately, we conclude that SN models do not have a dramatic impact to the method proposed in this paper. In particular, we have shown that $5\sigma$ statement can be made by using up to $\mathcal{O}(20)$ events.

\bibliographystyle{JHEP}
\bibliography{refs}

\end{document}